# Data Driven Prognosis: A multi-physics approach verified via balloon burst experiment

*Abhijit Chandra[1] and Oliva Kar[2]*
[1]Mechanical Engineering
[2]Computer Science and Human Computer Interaction
Iowa State University, Ames, Iowa
Contact: achandra@iastate.edu

*Abstract* — A multi-physics formulation for Data Driven Prognosis (DDP) is developed. Unlike traditional predictive strategies that require controlled off-line measurements or "training" for determination of constitutive parameters to derive the transitional statistics, the proposed DDP algorithm relies solely on in-situ measurements. It utilizes a deterministic mechanics framework, but the stochastic nature of the solution arises naturally from the underlying assumptions regarding the order of the conservation potential as well as the number of dimensions involved. The proposed DDP scheme is capable of predicting onset of instabilities. Since the need for off-line testing (or training) is obviated, it can be easily implemented for systems where such a priori testing is difficult or even impossible to conduct. The prognosis capability is demonstrated here via a balloon burst experiment where the instability is predicted utilizing only on-line visual observations. The DDP scheme never failed to predict the incipient failure, and no false positives were issued. The DDP algorithm is applicable to others types of datasets. Time horizons of DDP predictions can be adjusted by using memory over different time windows. Thus, a big dataset can be parsed in time to make a range of predictions over varying time horizons.

## I. INTRODUCTION

Predictive or prognostic capability represents a core competency required in many walks of life. Traditionally, analytical or numerical models based on conservation principles are used to make such predictions. These models require satisfaction of three principles: (1) equilibrium (which embodies satisfaction of a conservation principle), (2) compatibility, and (3) constitutive relations. In most cases, the constitutive relations depend on prior knowledge of constitutive or material parameters. Most often, a failure criterion and associated critical or threshold value of a parameter (typically a scalar) is also required to predict onset of instabilities. Such estimation protocols require controlled off-line testing or "training" for the prediction model. Further, applicability of such a prediction model is limited to situations spanned by the "column space" of the training set.

Numerous applications of practical interest also suffer from two major drawbacks: (1) the explicit expression for conservation principle or utility function being conserved may be unknown, (2) the system may not be amenable to perform off-line (in particular, destructive) testing. For instance while trying to predict the onset of instability in a material (e.g. necking in a tension test done on a material), the material properties can be estimated even before the actual experiment is done. The inferred constitutive parameters are then utilized for prognosis during the actual experiment. But if the part under consideration is already in service (e.g., in a structure or part of a person's anatomy), it may not be accessible for such off-line parameter evaluation. This is also true when a non-materialistic or abstract system (e.g. a global economic system, genetics or a health care system) is considered, there is neither any way to know the specific system properties, nor there is any method to assess the explicit nature of the conservation potential. In such cases, there is a need to rely on a predictive algorithm that can function without a priori knowledge of these parameters.

The present work circumvents these two difficulties by devising a DDP algorithm. The proposed algorithm requires assumption of the existence of a conservation principle, but its exact form need not be specified a priori. Instead, the exact form of the conservation functional is only specified locally, in the neighborhood of each observation point, (and is assumed to be piecewise quadratic in the current work) while its global form remains unknown. The conservation principle is then defined as the minimization of system curvature at each of the observation points. Based on such minimization principle, a dimensionless length scale of the underlying phenomenon is estimated at each observation point in each individual dimension. The constitutive parameters of any system (physical or abstract) are contained (with correct combination rule) within the expression for dimensionless length scales. Finally, stability is defined as the ability of the system to minimize its local curvature below a threshold value (determined as inverse of length scale in the present work) at each individual observation points. Thus, the stability characteristic of a system at a given location depends only on the local curvature of the system and the instantaneous dimensionless length scale at that location, obviating the need for explicit estimation of constitutive parameters.

## II. LITERATURE REVIEW

The key to useful prognosis is consistent estimation of future trend based on past experience and current observations defining the current state of the system. Two key questions are: (1) Will the current trend continue over a pre-specified prediction horizon?

Data Driven Prognosis: A multi-physics approach

And (2) What is the remaining life or time-span over which current activities can be continued? A detailed literature review along these lines may be found in [1]

Prognostics and health management (PHM) capabilities combine sensing and interpretation of data on the environmental, operational, and performance-related parameters of the product to assess the health of the product and then predict remaining useful life (RUL). This data is often collected in real-time or near real-time and used in conjunction with prediction models to provide an estimate of its state-of-health or degradation and the projection of remaining life. Traditionally, these prediction models use either a model based approach or a data-driven approach.

Model based approaches rely on an understanding of the physical processes and interrelationships among the different components or subsystems of a product [2], including system modeling and physics-of-failure (PoF) modeling approaches.

In system modeling approaches, mathematical functions or mappings, such as differential equations, are used to represent the product. The constitutive parameters or coefficients of the differential operators must be known a priori or estimated off-line based on a controlled set of stimulations. Statistical estimation techniques based on residuals and parity relations are then used to detect, isolate, and predict degradation [2-3]. Model-based prognostic methods have been developed for digital electronics components and systems such as lithium ion batteries [4], microprocessors in avionics [5], global positioning systems [6], and switched mode power supplies [7].

One of the examples of statistical estimation techniques and system modeling that is utilized for anomaly detection, prediction and diagnosis is the "divide-and-conquer" [2] dynamic modeling paradigm. The system input–output operation space is partitioned into small regions using self-organizing maps (SOMs) and then a statistical model of the system expected behavior within each region is constructed based on time–frequency distribution (TFD).The significant deviations from the trained normal behavior are recognized as anomalies. Then "diagnosers" can be constructed for various known faults within each operational region to identify the types of faults. This divide-and-conquer approach leads to a localized decision-making scheme, where anomaly detection and fault diagnosis can be performed locally within each operational region.

Another system modeling prognostic method used for diagnosis for lithium ion batteries is the Bayesian Framework approach. [4] The Bayesian learning framework attempts to explicitly incorporate and propagate uncertainty in battery aging models. The relevance vector machine (RVM)–particle filter (PF) approach provides a probability density function (PDF) for the end-of-life (EOL) of the battery. RVM is a Bayesian form representing a generalized linear model of identical functional form of the Support Vector Machine (SVM). Both of the above methods use some sort of supervised learning.

PoF based prognostic methods utilize knowledge of a product's life cycle loading conditions, geometry, material properties, and failure mechanisms to estimate its RUL [8-11]. PoF methodology is based on the identification of potential failure mechanisms and failure sites of a product. A failure mechanism is described by the relationship between the in situ monitored stresses and variability at potential failure sites. PoF based prognostics permit the assessment and prediction of a product's reliability under its actual application conditions. It integrates in situ monitored data from sensor systems with models that enable identification of the deviation or degradation of a product from an expected normal condition and the prediction of the future state of reliability. Such methodology requires establishment of "benchmark" parameters or normal behavior of the product. This requires controlled off-line testing or controlled "training" regimen.

PoF based approach has been applied to analyze the health of printed circuit boards (PCB) to vibration loading in terms of bending curvature [8]. First, the components which are most likely to fail and their locations are identified at certain vibration loading levels. Sensors are placed at those areas to monitor the PCB response. Then a database is built that reflects the relation between the PCB and its critical components. Similar approaches have also been applied for surface mount assemblies to extract the state of damage at an instant and predict the RUL based on accumulated damage [12].

The data-driven approach uses statistical pattern recognition and machine learning to detect changes in parameter data, isolate faults, and estimate the RUL of a product [13-16]. Data-driven methods do not require product-specific knowledge of such things as material properties, constructions, and failure mechanisms. In data-driven approaches, in-situ monitoring of environmental and operational parameters of the product is carried out, and the complex relationships and trends available in the data can be captured without the need for specific failure models. There are many data-driven approaches, such as neural networks (NNs), SVMs, decision tree classifiers, principle component analysis (PCA), PF, and fuzzy logic [13]. However, such techniques also require a definition of normal operation, which is typically based on a training set or previously observed circumstances. Thus, current state-of-the-art predictions in data driven approaches only extend to circumstances that can be spanned by the 'column space' of the training conditions. This severely limits their usefulness, particularly in unforeseen circumstances. Alternatively, developing a data driven scheme for a specific purpose requires very carefully articulated training regimen (capturing the purpose) for the prognostic system.

Neural networks are examples of traditional data driven systems, where the data processing is carried out at a number of interconnected processing elements called neurons. The neurons are usually organized in a sequence of layers, an input layer, a set of intermediate layers, and an output layer. During training, the network weights are adjusted depending on the type of learning process. A set of feed forward back propagation networks are used that undergo supervised learning to identify the current operating time of an operating bearing. Two classes of neural network models, single-bearing models and clustered-bearing models have been developed. [17] Both classes of models use degradation information associated with the defective phase of bearing degradation. In first class, a single bearing is used to train a single back propagation neural network. In the second class, the bearings are classified in groups (clusters) based on similarity in their failure and defect times. Each net is then trained using degradation information associated with bearings in the cluster. [17]

Thus, current state-of-the-art in both system modeling and data driven approaches is limited by the prior training requirement. Controlled off-line or a priori testing is needed for accurate constitutive parameter estimation. Moreover, the applicability of a prognostic system is strictly limited to the vector space spanned by the basis vectors of its training regimen.

By contrast, the proposed DDP approach estimates the relevant constitutive parameters in-situ. Instead of carefully designed "training" regimen, it can utilize any two time sequences of data. However, it only focuses on the situation at hand, rather than trying to master any general situation. Accordingly, only a relevant combination of material and geometric parameters is extracted as an instantaneous dimensionless length scale in each dimension, and utilized in real time. Such a DDP scheme is then capable of predicting the probability of the onset of instability over a prediction horizon. Alternatively, the prediction horizon needed for the probability of instability to exceed a threshold value can be considered the RUL.

## III. DETERMINATION OF LENGTH SCALE

Let us consider an observable body or a phenomenon containing a finite number of observation points. At each point, information is collected at multiple dimensions (that collectively satisfy work conjugacy requirements), and at discrete instants of time. Let us now consider two specific observation points A and B. At both of these points, each dimension is denoted by $i = 1, n$. The value recorded at these two points in some particular dimension $i = d$ can be denoted as $u_d^A$ and $u_d^B$. For the present analysis, after all values are recorded at every observation point at a time instant, a normalized relationship is developed between each pair of points in each dimension. Such a normalized relationship between two points A and B is given by $a_d^{AB} = \frac{u_d^A - u_d^B}{u_d^A + u_d^B + 2\bar{m}}$, where $\bar{m}$ is an arbitrary small constant that is determined later.

In order to develop a model describing such a phenomenon, it is first assumed that the system under observation is conservative. As a first attempt, it is also assumed that a piecewise second order potential is sufficient to describe the pairwise interactions in the system. Thus, the general potential function is assumed to be quadratic in the neighborhood of each observation points. However, the nature of the quadratic potential function (coefficients of the second order polynomial) can vary from one point to the other, representing a higher order global relationship. The approximations and intrinsic uncertainty introduced in the model predictions due to this approximation will be examined later.

Next, it is attempted to satisfy the three canonical requirements: (1) compatibility, (2) equilibrium and (3) constitutive relation. It is further acknowledged that objectivity or frame invariance (with respect to the observer) is a requirement for describing the behavior of such a system. Objectivity implies that the state of the observed system remains invariant with respect to different observers or variations in the observation procedure. In the present development, compatibility is enforced indirectly by requiring that the system be objective at every observable scale at each location [18]. Such an indirect approach satisfies compatibility as well as objectivity simultaneously. Using such an approach [18], the conservation of linear momentum in the neighborhood of point A may be described as,

$$R_i^A - \beta_{ik}^A * \Delta H_k^A = 0 \qquad (1)$$

Here $R_i^A$ is the rank at point A in dimension $i$ [18]. Such a rank satisfies both compatibility (and objectivity) as well as equilibrium at A. $H_k^A$ represents the Borda Count [19-23], and $\Delta H_k^A$ represents the change in the Borda count at point A in dimension $k$ during a time step. The parameter $\beta_{ik}^A$ may be described as $\beta_{ik}^A = \frac{1}{(\Delta t)^2} * \frac{\rho}{E_{ijkl}} * \|L_l L_j\|$. $\rho$ is density and $E_{ijkl}$ is tangent modulus. The parameter $\beta_{ik}^A$ is a non-dimensional quantity and essentially represents a second order norm of the length-scale around the point A, in which the linearized (1) is valid. The length-scale essentially denotes a region around the observation point, in which the piecewise quadratic assumption for the potential function is valid. So substituting $\beta_{ik}^A$ in the equation above, we obtain an expression for length scale $(L_l)$:

$$R_i^A - \frac{1}{(\Delta t)^2} * \frac{\rho}{E_{ijkl}} * L_l L_j * \Delta H_k^A = 0 \qquad (2)$$

$$\Rightarrow R_i^A = \frac{1}{(\Delta t)^2} * \frac{\rho}{E_{ijkl}} * L_l L_j * \Delta H_k^A \qquad (3)$$

$$\Rightarrow E_{ijkl} \left[\frac{R_i}{L_l L_j}\right] = \frac{\rho}{(\Delta t)^2} * \Delta H_k \qquad (4)$$

Out of all the possible transformation laws, the ones that are permissible in the above (4) are the ones that conserve angular momentum and the symmetry of the potential function. Now the conservation of angular momentum requires that ($i$ and $j$) be interchangeable. Similarly, the interchangeability of ($k$ and $l$) is mandated by the symmetry requirements on the definition of strain. The requirement of work conjugacy necessitates symmetry in potential function and this enforces interchangeability between $(i\,;\,j)\ pair$ and $(k\,;\,l)pair$. After all such transformations, the above (4) can be written as:-

$$\frac{E_{ijkl}}{2} * \left[\frac{R_i}{L_l L_j} + \frac{R_j}{L_k L_i} + \frac{R_k}{L_j L_l} + \frac{R_l}{L_i L_k}\right] = \frac{\rho}{(\Delta t)^2} * [\Delta H_i + \Delta H_j + \Delta H_k + \Delta H_l] \qquad (5)$$

Since the number of solutions for the quadratic equation in $L_i$ is $2^{dimSize}$, 8 and 16 values of $L_i$ are obtained when number of total dimensions are 3 and 4 respectively. The number of possible solutions is called the number of roots of the length scale. It is interesting to note that solution for a dimensionless form $\bar{L}$ of the length scale containing the correct combination of geometric and material parameters is sufficient for our intended prognosis.

A constant $m_i^{AB}$ is evaluated [24] for each observation pair by assuming an order (quadratic in present work) of the interaction potential between the pair under consideration. Note, this order need not be the same for all pairs. Next, the constant $m_i^{AB}$ is evaluated by setting the spatial gradient of the observed variable at observation point A $(u_{i,j}^A)$ to be exactly the same as the spatial gradient of the normalized variable $(a_{i,j}^A)$. Finally, the constant $\overline{m}$ is evaluated from a least square fit of all the $m_i^{AB}$ values. The constant $\overline{m}$ essentially sets the datum, and the coordinate of the origin is set at $-\overline{m}$ in the respective dimension. Since only a least square approximation is used for evaluating $\overline{m}$, and it is used universally at all points (to set same datum for all observation points), the gradient of the observed variable and that of the normalized variable are not exactly the same at all points. This introduces an approximation in our formulation, and decides the limiting resolution for the current DDP methodology. The error introduced due to this approximation is measured in the present work.

### A. Energy Exchange Rate and Its Manifestation as a Curvature

It is assumed that equilibrium in the system is satisfied instantly, while satisfaction of compatibility (in an objective framework) only happens with time. Thus, at every instant, the system state adapts in an attempt to satisfy compatibility in addition to equilibrium. As a result, the Borda Count at an observation point A changes over the associated length scale (which itself can also change as a result of this adaptation). The local curvature reflects this change, which is manifested as a local energy exchange (absorption or release) rate. We assume that loss of stability occurs locally when the curvature at a point A exceeds a threshold value (denoted by inverse of dimensionless length scale). It is further assumed that global instability occurs when such unstable points can form a "chain" or an energy exchange pathway, and it meets two additional conditions: (i) the chain length exceeds a critical threshold, and (ii) the energy exchange rate along such a pathway exceeds a critical threshold. [24]

## IV. ALGORITHMS

The algorithms needed for evaluation of length scales, curvatures as well as instability prediction criteria are discussed in this section [24].

### A. Length Scale Estimation

This calculation follows a generic approach. This approach is followed so that the algorithm can be applied for any system where the actual Poisson's ratio of the system (or mode mixity between dialatational and distortional modes) is not known a priori. It is already known that Poisson's Ratio for pure volumetric deformation or dilatation $(\nu_v)$ = -1 and Poisson's Ratio for only shape change (at constant volume) or pure Shear $(\nu_s)$ = 0.5. Using these two prior known values we calculate the proportion of volumetric deformation and shear deformation in the phenomenon under observation by requiring that the combination minimizes the resulting curvature.

### B. Nature of Roots

At each point there are a number of dimensions. So, for one point there are $2^{dimSize} * dimSize$ number of roots of $\bar{L}$. . In each specific dimension there are $2^{dimSize}$ number of roots. Thus, only a stochastic estimate of the length scale at an observation point is possible. This introduces the stochastic characteristics of the resulting prognosis. In the present work, four dimensions (three spatial dimensions and grayscale color) are used. Thus, we obtain 16 roots at each observation point at every time step. This multiplicity in the number of roots gives rise to the stochastic nature of the prognosis protocol, and a probability for instability (over the specified time horizon) can be estimated simply by knowing what fraction of the roots are indicating incipient instability.

### C. Curvature Calculation

This calculation also follows a generic approach. It is assumed that, in any combination, equilibrium is satisfied instantly, while satisfaction of compatibility in addition to equilibrium takes longer time. Thus an original gradient of $\frac{H}{\bar{L}}$ satisfying equilibrium only, gradually transforms to $\frac{R}{\widetilde{\bar{L}}}$ (where $R$ is the objective rank [17] and $\widetilde{\bar{L}}$ is the transformed dimensionless length scale). Curvature of the system arises naturally due to such change in gradient in the neighborhood of an observation point. Pursuing an updated Lagrangian approach, the Composite Curvature $\kappa_i$ is calculated as $\frac{(R*\bar{L}_i - H*\widetilde{\bar{L}_i})}{(((\bar{L}_i{}^2)*\widetilde{\bar{L}_i})*(1+(H/\bar{L}_i)^2)^{1.5})}$ for each point, specific dimension $i$ and a specific root of length scale, and applied to all points.

*D.  Post Processing: Categorization of Observation Points at an Instant*

Following the curvature and associated length scale calculations, a Path Dependency Index (PDI) is calculated at each observation point, and the points are categorized according to the PDI. There are eight possible categories that characterize the nature of instabilities that can arise at each point. Assignment of the first seven categories are based on PDI, and will be explained in this section. The 8th & 9th categories are assigned after further processing. It is based on the Global Transcendation Index (GTI) which will be explained in the next section. These categories are summarized in Table 1.

*E.  Chain Length Calculation*

The chain length determination is done at every point for each dimension and each root [24]. This is done to check how long does a defect continue in either directions from one point to another point in order of their ranks. Proximity is defined by difference in Rank. The curvature combination rules of Saari [25] are used, and it is assumed that sufficient time is allowed to reach steady state after each combination. In addition to material systems (e.g., balloon burst phenomenon), this facilitates chain length computation in abstract systems (e.g., Economic systems) without specific proximity definitions. These chains constitute energy exchange pathways in the system, and to be on an energy exchange pathway, a point must have a PDI >=5. A chain length that reaches critical threshold is assumed to transcend to the next aggregated scale in hierarchy, and is "promoted" to show up as a "local" instability at the next level of aggregation.

*F.  Global Transcendation Index*

After the system attains path dependency (at a local level) it may progress towards instability and failure if two additional criteria are also met: (i) The locally path dependent points link in forming a chain, whose length exceeds a threshold value, and (ii) The aggregated or "residual" curvature" of the entire system exceeds a critical threshold. The Global Transcendation Index (GTI) turns nonzero and positive if both of these conditions are met. A category of 8 is assigned for an observation point if PDI >5 and GTI >0 in one dimension, and a category of 9 is assigned if both of these conditions are met in multiple dimensions.

This section describes the "zoom out" or aggregation procedure for calculating the critical chain length and the residual (or aggregated) curvature for the system at a time instant.  The residual curvature provides a measure of the energy exchange rate of the system as a whole with its environment. Further details may be found in [24].

The "zoom out" procedure utilizes the fact that for a truly conservative and isolated system, the observed curvature should be zero when the "stand-off" distance of the observer is both zero and infinity. The x-axis in Fig. 1 represents such stand-off distance or "zoom-out" level, while y-axis represents the system curvatures estimated under such situations. A log-scale in x is used for convenience. It is assumed that at x-value of zero,  the system curvature drops to zero. However, due to resolution limitations of the imaging system, it was not possible to image at that level. Thus, the level of aggregation of the most detailed acquired image was arbitrarily assigned a value of unity, and only "zoom-out" was carried out with increasing levels of aggregation. The residual curvature value when further zoom-out could not be carried out is assumed to be a x-value of infinity. However, in practice, it represented a level when the entire video frame was aggregated to 9 (3x3) points. The curvature value at this aggregated extremum is called the "Residual Curvature" of the system. A non-zero value of residual curvature represents a measure of the energy exchange rate through the system. Due to the nature of our calculation, only a magnitude of the residual curvature is obtained. It is both positive and negative, simultaneously. This is due to the conservation assumption in our analysis, that requires any energy absorbed (by the system as a whole) to be released within the time step under consideration, and vice versa.

Since, we do not have any physical data more detailed than the captured pixel level in our image, we arbitrarily assume that the plot is symmetric to the left and right of x=1 and drops to zero at x=0.

We calculate $\kappa, \frac{1}{LTilda}, \frac{1}{L}$ at each aggregation levels. $\frac{1}{L}$ is denoted as "Kappa-short" and $\frac{1}{LTilda}$ is denoted as "Kappa-long", since these are considered short term and long term critical values of local curvature.

The aggregation level represents the number of points that were considered together as a unit for the specific calculation. From Fig. 1, It is attempted to extrapolate the Kappa graph backwards for lower values of x extending to zero by using the assumed symmetry condition.After this the $\frac{1}{LTilda}$ line is extended backwards to intersect the mirror image (about y-axis) of the calculated Kappa line..The existence of a process zone associated with local instability is assumed. It is further assumed (Fig.1) that the process zone size instantly jumps to point B if it reaches point A, provided the y-value (or curvature) at B is lower than A. The log(x) value at intersection point A is converted to fractional number of points relevant to the aggregated frame under consideration, and is scaled by the actual frame under consideration to reach an estimate for the critical chain length.  The intersection of $\frac{1}{LTilda}$ line provides an estimate for the long term critical chain length while intersection with the $\frac{1}{L}$ line provides a short term estimation of the critical chain length.

Data Driven Prognosis: A multi-physics approach

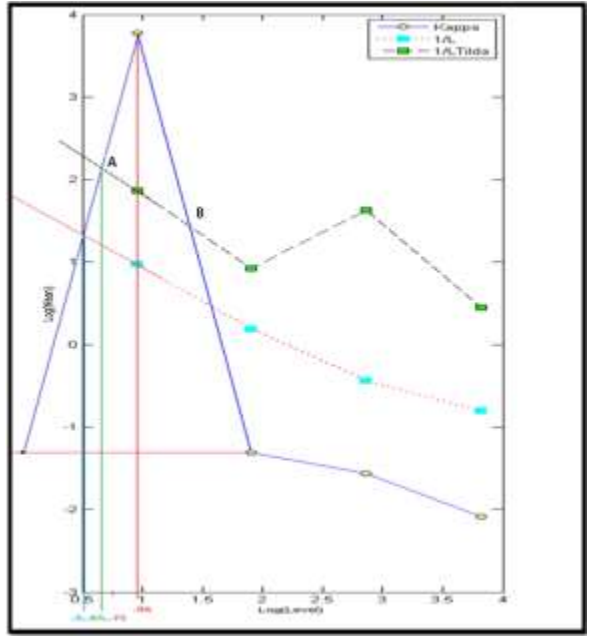

*Figure 1: Critical localization index calculation for Balloon Burst (online version in color)*

V. BALLOON BURST EXPERIMENT

*A. Data Collection*

    The balloon bust experiment consisted of mechanically blowing a balloon until it "popped" or failed by bursting. The XYZM files that provided length, breadth, depth and color information of the balloon as a 3-D object are collected as a sequence of gray scale video frames over the entire duration of the experiment.

*B. Experimental Set Up*

    The 3-D balloon video (Fig. 2 shows sample video frames) is taken as a sequence of high-speed real time 3-D shape measurements based on rapid phase-shifting technique [26]. The experimental system takes full advantage of the single-chip DLP (Digital Light Processing) technology for rapid switching of three coded fringe patterns. A color fringe pattern with its red, green, and blue channels coded with three different patterns is created by a personal computer. When this pattern is sent to a single-chip DLP projector, the projector projects the three color channels in sequence repeatedly and rapidly. To eliminate the effect of color, the color filters on the color wheel of the projector are removed. As a result, the projected fringe patterns are all in grayscale. A properly synchronized high-speed black-and-white (B/W) CCD camera is used to capture the images of each color channel from which 3-D information of the object surface is retrieved. A color CCD camera, which is synchronized with the projector and aligned with the B/W camera, is also used to take 2-D color pictures of the object at a frame rate of 26.7 frames/sec for texture mapping. Along with this system, a fast 3-D reconstruction algorithm and parallel processing software [26] is also used to realize high-resolution, real-time 3-D shape measurement at a frame rate of up to 40 frames/sec and a resolution of 532x500 points per frame. The XYZM files are just the XYZ points that get triangulated (one for every pixel of the 2D image) and then the BMP file is the texture which is simply the three captured images averaged together [26].

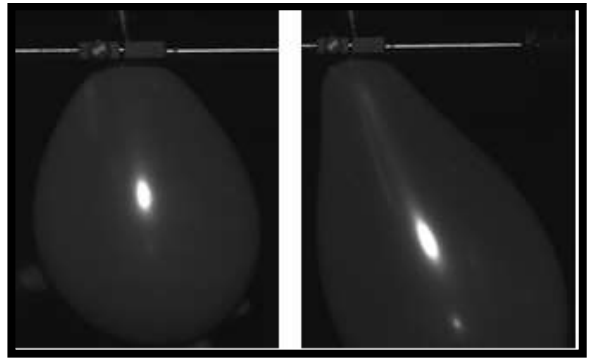

*Figure 2: Balloon in the beginning and near the end before bursting*

## C. Collected Data Files

The input files are in "xyzm" format. Each of the observation point represents a pixel of the video captured of the balloon being blown. Each observation point has 4 dimensions or 4 types of information (x, y, z and mask information). From these 4 types of information, the x – coordinate, y – coordinate, z – coordinate and color information for each of the observation point is obtained. These 4 kinds of information form the four dimensions that the prediction algorithm uses. These four dimensions together form a conservative system. The value of the color information decreases as air is pumped into the balloon while the value of the x, y, z information increases. Since, it is a video file, the collected data can grow very quickly in size. A 5 minute video (at 40 Hz) will typically result in almost 3200 GB of data for XYZM type of files. The proposed DDP algorithm reduces memory requirements by parsing such data, and utilizing only frames at n∆t (n = 1, 2, 3, etc.) intervals. Thus, past memory over only n∆t is needed at any time instant. The proposed DDP algorithm can function very well with n = 1, while higher values of n extends the prediction horizon. Of course, the implicit assumption in the prognosis scheme remains that the dimensionless length scales calculated over previous n∆t remains valid over future n∆t steps. This limits the upper bound of n or the prediction horizon that can be used at any time.

## VI. MODEL BASED PROGNOSIS AND VERIFICATION

### A. Path Dependency Index

Since the data size for the video file grows very rapidly, the first objective of our analysis is to make a prognosis using current observation, and hold a minimal number of frames in memory. As a first attempt, only two frames (current frame and its immediate predecessor) are used at each time instant.

First, the local Borda Count and Objective Rank are calculated at each of the observation point, at each instant of time. Using two consecutive time instances, the dimensionless length scales are calculated at each observation point. This allows calculation of the local curvature at each observation point. The local curvature at a time step is compared to the threshold value or critical curvature (denoted by inverse of length scale), and a Path Dependency Index (PDI) is calculated (Table 1). A PDI greater than or equal to 5 indicates a local instability with potential to transcend to global scale.

Next, we follow an objective aggregation procedure consistent with the assumption of piecewise and pairwise second order potential [24]. This is called the "Zoom Out" procedure that identifies the critical "chain length" or length of the energy exchange pathway needed for local instabilities to transcend to global scales.

However, the existence of a chain length greater than the minimum or critical length only constitutes a necessary condition for such transcendation. The energy exchange rate through such a pathway must also exceed a threshold value to meet the sufficiency condition for transcendation of local instabilities to a global scale. The critical energy release rate is a constitutive property that can be normally determined accurately only with careful off-line testing. Instead of such off-line testing, we utilize the fact that whenever local instabilities transition to global scales, almost all of the energy stored along such a pathway gets released (transformed to another form). Thus, the energy exchange rate rises, and then falls very rapidly to near zero during such transcendation phenomena. We utilize this rapid change in energy release rate in our prognosis scheme, and a trigger is initiated whenever the dimensionless energy release rate drops by more than 80% within a single time step. The choice of 80% is arbitrary in this case, based on inherent noise floor in our experimental and computational procedures.

Together, the existence of: (i) greater than critical chain length, and (ii) greater than 80% drop in energy release rate over a single time step, constitute a positive reading for the GTI, and GTI is set >0.

We continue the calculation of PDI and GTI for every time step of acquired data to prognosticate about the balloon burst phenomenon. When PDI >5 and GTI >0, it indicates the imminent bursting of the balloon. The model based prognosis results are finally compared with the actual time of balloon burst that are experimentally observed.

Progression of Path Dependency Index (PDI) for the color dimension across all times is shown in Fig. 3. It shows the percentage of points that have reached different PDI markers as the balloon gets blown bigger with progression of time. The different PDI categories are explained in Table 1. Fig. 4 shows a plot that signifies the instants when the system shows local instabilities (PDI >=5), without regard to the chain length information. For a point to be counted as path dependent, at least 8 of the 16 roots must show a PDI >=5. If at least one point is path dependent, the system is considered path dependent. So for the balloon, path dependency started at time index 4 and the balloon is also path dependent at time indices 13, 16, 22.

Fig. 5 shows the number of roots (out of 16) in color dimension that had PDI >=5, and also crossed the critical chain length threshold (27 for the balloon experiment in color dimension [24]). From Fig. 5 the balloon has crossed the critical chain length at time indices 20, 21, 23, 24 and 25.

TABLE 1. PATH DEPENDEMCY INDEX CATEGORIES

| Path Dependency Index Categories | |
|---|---|
| *Categories* | *Meaning of Categories* |

Data Driven Prognosis: A multi-physics approach

| Category 1 | Full Stability : $\frac{abs(Kappa)}{abs(KappaLong)} < 1$ and $\frac{abs(Kappa)}{abs(KappaShort)} < 1$ |
|---|---|
| Category 2 | Short term instability but long term stability: $\frac{abs(Kappa)}{abs(KappaShort)} > 1$ |
| Category 3 | Short term stability but long term instability: $\frac{abs(Kappa)}{abs(KappaLong)} > 1$ |
| Category 4 | Conditional Stability: Possibility exists for controlling local instability by altering mode mixity of dialational and distortion modes. |
| Category 5 | Both short term and long term instability. Control by alteration of mode mixity is not possible. Instability in single dimension. |
| Category 6 | Both short and long term instability in more than one dimension. |
| Category 7 | Both short and long term instability in more than two dimensions |
| Category 8 | PDI>5, Energy Exchange Channel(s) formed with Chain Length > short term critical chain length, |
| Category 9 | PDI>5, Energy Exchange Channel(s) formed with Chain Length > long term critical chain length |

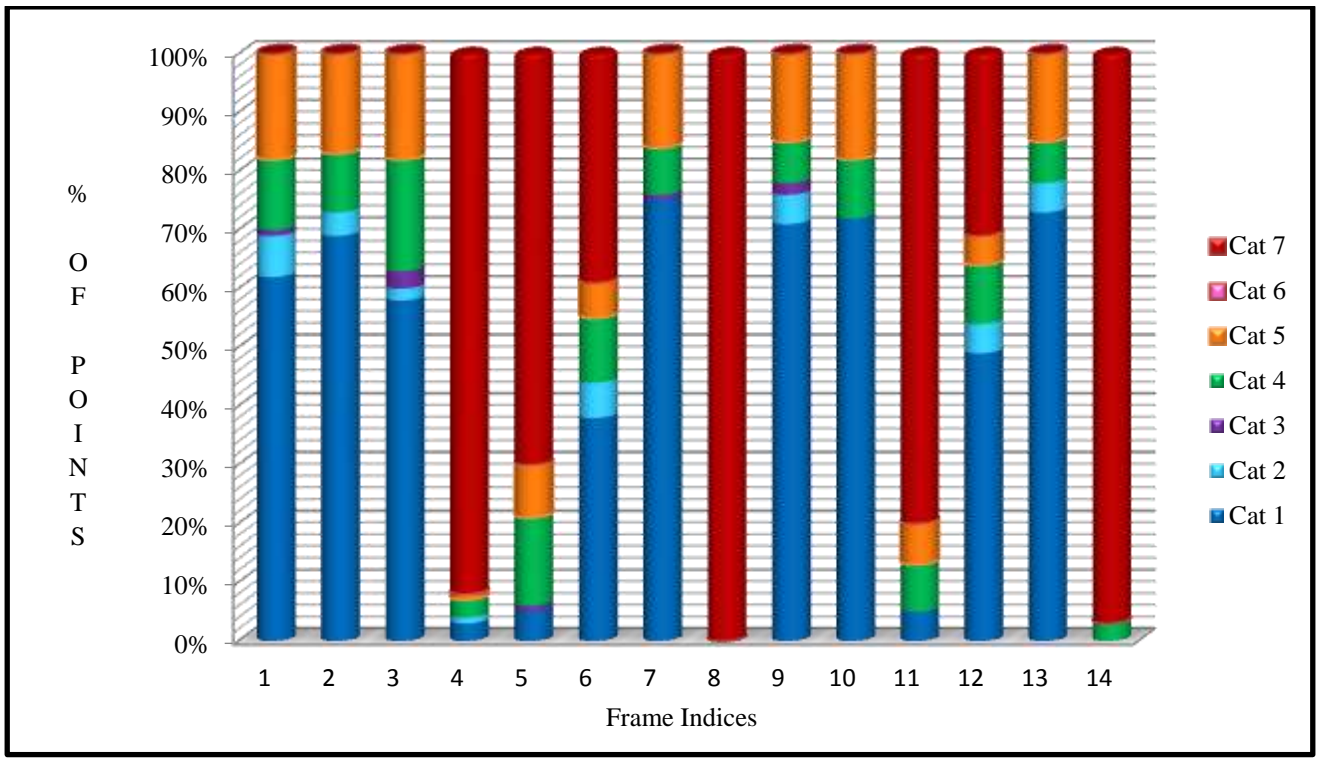

*Figure 3: Progression of Path Dependency Index with time (in Color Dimension) (online version in color)*

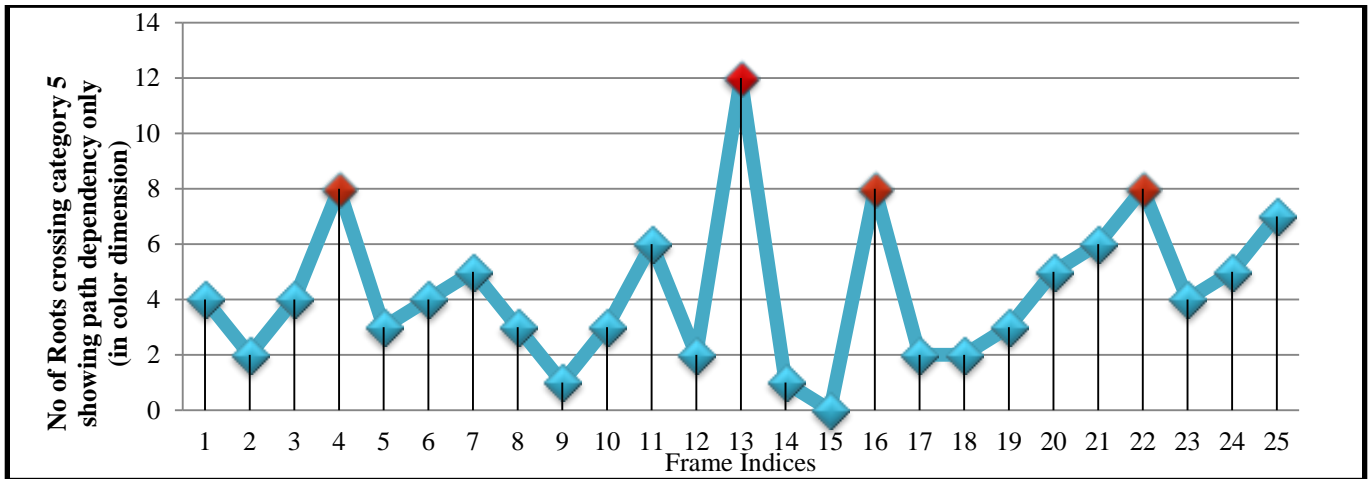

Figure 4: Number of roots having category greater than 5 showing path dependencies (in Color Dimension) (online version in color)

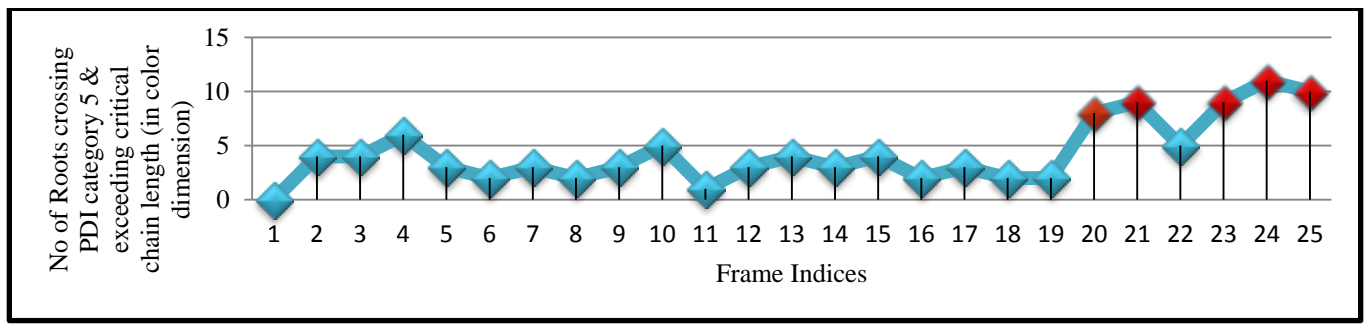

Figure 5: Number of roots having category greater than 5 and having chain length greater than critical number of points (online version in color)

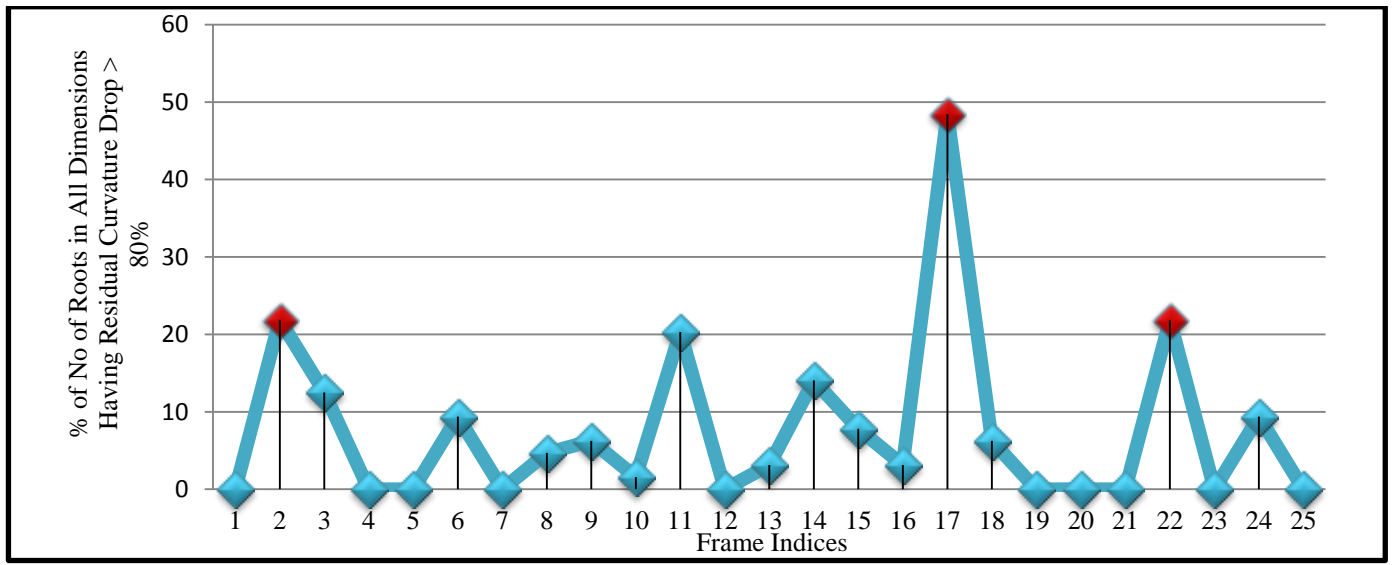

Figure 6: Number of roots having residual curvature drops more than 80% (online version in color)

*B. Residual Curvature*

Fig. 4 explains when the system is path dependent whereas Fig. 5 explains when there is a possibility for the local instability to transcend to global scales by forming energy exchange channels longer than the critical threshold. Fig. 6 shows residual curvature for the system at different time indices. Residual curvature provides a dimensionless aggregated measure of the magnitude of the energy exchange rate for the entire system, and a local instability can transcend to global scale if the residual curvature also exceeds a critical threshold. However, such threshold can only be measured via carefully controlled off-line testing. Here, we utilize another fact that the stored energy gets rapidly depleted during any such global transcendation phenomenon. Hence, to detect imminent balloon burst, we monitor the rate of change, particularly a rapid drop from an elevated value for the residual curvature. A trigger is initiated whenever the residual curvature drops by more than 80% within a single time step. Fig. 6 shows the percentage of total number of roots in all four dimensions of the balloon (out of 64 roots) that had a drop of residual curvature greater than 80%. The instants of time where the number of roots reached a value greater than 20% (approximately 13 roots out of 64) are said to be the time instants where the balloon is dissipating large amount of energy. The time indices where this happens are 2, 17 and 22.

## C. Composite Failure Prediction

The final failure prediction takes all of the above analysis into account. For a system level failure to occur, first the system needs to enter the path dependent stage. A system can be stable even if it is path dependent. While being path dependent, the system can nucleate "dislocations" at multiple points. The points which are only contiguous to each other in rank can actually form a chain. If the chain formed in such a manner exceeds the short term or long term critical chain length of the system then the dislocations can join and form a line defect which may ultimately give rise to failure. In addition to this, the system needs to have greater than critical energy exchange rate. This activity is monitored via a large and rapid drop in residual curvature. A drop of greater than 80% is used as a trigger in our analysis. Once both of these take place (GTI >0) after the system has already entered the path dependency mode (PDI >5), failure can be predicted for that system. In the case of the balloon, the system was under constant monitoring when it entered the path dependency stage at time index 4. After time index 4, it had triggered long term chain formations at time indices 20, 21, 23, 24 and 25. It had also triggered residual curvature drops at time indices 17 and 22 after time index 4. So the time index by which all of the phenomena had taken place is 20. Hence the balloon was supposed to be approaching failure at that time index and the blow-up would be expected soon. Time index 20 corresponds to a frame number of 2099. In reality, the balloon had burst at frame 2382. Hence the DDP scheme predicts the failure $\left(1 - \frac{2099}{2382}\right) * 100 = 11.88\%$ ahead of actual time.

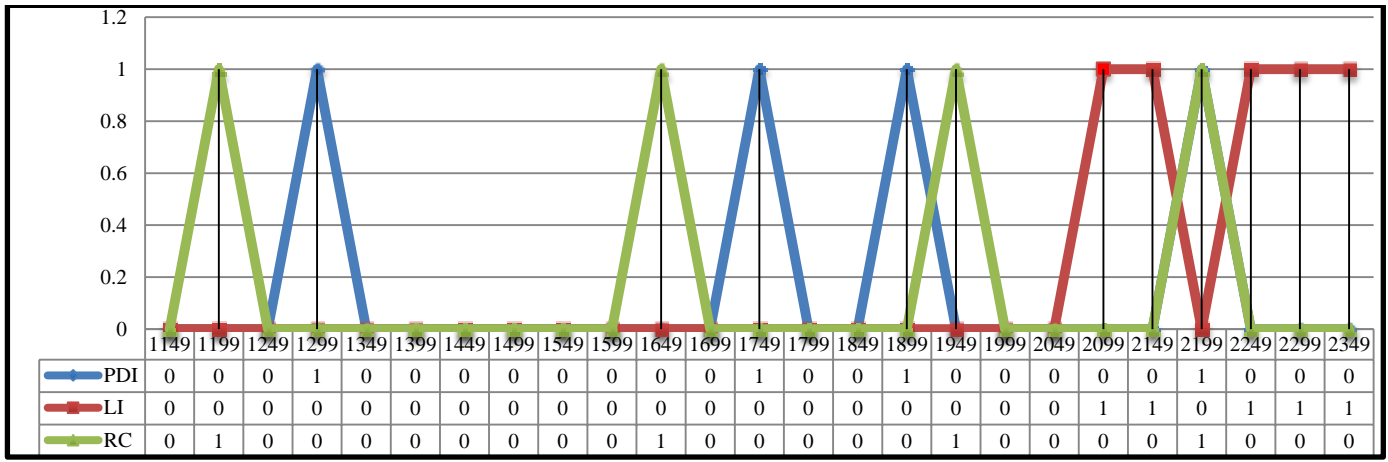

*Figure 6: Composite Failure Prediction of Sample Data Set 1 (Row 353-362, Col 343-352) (online version in color)*

## D. Results of other data sets for balloon verification

Similar analyses were also carried out using: (i) randomly selected different viewing windows on the same balloon, as well as (ii) different balloons. Tables 2 and 3 record the video frame indices pertaining to significant transitions in these additional datasets.

## VII. SUMMARY OF RESULTS FROM OTHER SETS

TABLE 2. RESULTS OF OTHER DATA SETS ON SAME BALLOON

| Dataset No. | Summary of Results | | | Composite Prediction Ahead % |
|---|---|---|---|---|
| | PDI started | Energy Dissipation | Chain Formation | |
| Set 1 | 1349 | 1649, 1949, 2199 | 2349 | 1.385 |
| Set 2 | 1599 | 1649, 1949, 2199 | 2349 | 1.385 |
| Set 3 | 1299 | 1649, 1949, 2199 | 2249 | 5.5835 |

TABLE 3. RESULTS FOR ADDITIONAL BALLOONS

Data Driven Prognosis: A multi-physics approach

| Dataset No. | Summary of Results | | | Composite Prediction Ahead % |
|---|---|---|---|---|
| | *PDI started* | *Energy Dissipation* | *Chain Formation* | |
| Balloon 3 | 301 | 601 | 351, 451 | 13.1503 |
| Balloon 4 | 451, 551, 601, 651 | 701 | 2349 | 24.1342 |
| Balloon 5 | 451 | 701 | 651 | 25.5048 |
| Balloon 6 | 451 | 601, 701 | 851 | 2.2962 |

## VIII. DISCUSSION AND CONCLUSION

A DDP algorithm suitable for handling multi-scale and multi-physics problems is developed here. The DDP algorithm is verified against balloon burst experiments using in-situ grayscale video data only, and no off-line testing. The DDP prognosticator never failed to predict an incipient instability. Thus, the failure predictions were conservative and always contained a safety margin.

Table 2 shows results for three datasets collected on a single balloon at different arbitrary locations. The DDP algorithm predicted occurrence of balloon burst approximately 5% ahead of actual occurrence observed experimentally. The worst case was a prediction about 12% ahead, and best case prediction was about 2% ahead.

Additional tests were performed with completely different sets of balloons (different brand name and size). While there is still no false positive, the DDP algorithm predicted failure much earlier for this additional set. The worst prediction is 25.5% ahead of actual failure, and the best prediction is 2.3% ahead. It has been observed that noise levels during these additional experiments used to collect data (Table 3) were significantly higher in the laboratory. Moreover, it is believed that the polymers of the additional balloons were different, and could be toughening during the test as a result of the imposed deformation. In our DDP protocol, the critical chain length from the zoom-out calculation was only estimated once near the beginning of the test. This was a direct consequence of attempts to reduce computational burden, since the zoom-out calculation was computationally intensive. Moreover, the critical chain length estimation procedure needed human intervention. This might have resulted in a smaller estimate of critical chain length compared to the toughened state of the underlying balloon polymer, and contributed to pre-mature failure predictions. We believe a protocol using re-calculation of critical chain length from the zoom-out (at every time step or least a periodic update) would have been better, and will be attempted in future.

A further approximation occurs in the zoom-out procedure due to the assumption that curvature is highest at the observed scale. This need not be strictly true, but can be rectified only by collecting data at several scales, and progressing with increased level of details (or zooming in) until a maxima in curvature occurs.

The reliance of the DDP prognosis on only short term memory (at least two data frames need to be held in memory at an instant) significantly reduces the requirement that long data sequences be held to infer transition statistics of the system. However, big data sets in spatial dimensions must still be handled. The proposed algorithm attempts to conquer such big data sets by organizing them in a hierarchy of scales. However, data can only be aggregated. This implies, the prognosticator can only traverse from local (or detailed) scale to aggregated or global scale, but not vice versa.

The proposed DDP algorithm obviates the need for a priori off-line testing or "training" to extract the transitional statistics of the system. Instead of attempting to develop a capability for predicting system response under general conditions, it utilizes only on-line available data and short term memory (minimum two time frames are needed) to develop a prognosticator specializing in the "current" situation or "problem at hand", and makes a prediction regarding incipient instability over a relatively short prediction horizon. Greater number of time frames may be used to stretch the prediction horizon. For remaining useful life (RUL) estimates, the best type of data for the proposed algorithm are those collected over short time windows separated longitudinally over a much larger time span. The separation time for the windows can be varied compared to the span of the data collection window to facilitate prognosis over varied prediction horizons. However, such a scheme requires capability for both short-term and long-term memory.

Current implementation of the proposed DDP algorithm is much slower than real time due to limitations of computational capability – both in terms of processor speed as well as number of processors deployed in parallel. The need for human intervention in critical chain length estimation also represents an obstacle against achieving real time implementation. Work is in progress to improve both of those aspects to approach real time implementation of the proposed DDP prognosticator.

The DDP algorithm developed here is also applicable to any datasets representing a conservative system. A companion paper applies similar framework for prognosis of osteo-arthritis after reconstructive surgery following ACL injury [27].

## ACKNOWLEDGMENTS

This work is supported by NSF through grant numbers: CMMI 0900093 and CMMI 1100066. The authors gratefully acknowledge this support. Any opinions, conclusions or recommendations expressed are those of the authors and do not necessarily reflect views of the sponsoring agencies.

Data Driven Prognosis: A multi-physics approach